\def\year{2020}\relax
\documentclass[letterpaper]{article} 
\usepackage{aaai20}  
\usepackage{times}  
\usepackage{helvet} 
\usepackage{courier}  
\usepackage[hyphens]{url}  
\usepackage{graphicx} 
\usepackage{subcaption}  
\usepackage{graphicx}  
\title{Synthesizing Credit Card Transactions}

\author{Erik R. Altman\\
IBM T.J. Watson Research Center\\
1101 Kitchawan Road\\
Yorktown Heights, NY  10598\\
ealtman@us.ibm.com
}

\begin{document}

\maketitle

\begin{abstract}

As noted by Turing Laureates Geoffrey Hinton and Yan
LeCun~\cite{2019-Turing-Lecture}, two elements have been essential to AI's
recent boom:  (1) deep neural nets and the theory and practice behind them;
and (2) cloud computing with its abundant labeled data and large computing
resources.

Abundant labeled data is available for key domains such as images, speech,
natural language processing, and recommendation engines.  However, there are
many other domains where such data is not available, or access to it is
highly restricted for privacy reasons, as with health and financial data.
Even when abundant data is available, it is often not labeled.  Doing such
labeling is labor-intensive and non-scalable.

To get around these data problems there have been many proposals to generate
synthetic
data~\cite{Rubin93,Peng15,Patki16,Li17,Neuromation-Synth-Data,Capital-One-Synth-Data}.
However, to the best of our knowledge, key domains still lack labeled data or
have at most toy data; or the synthetic data must have access to real data
from which it can mimic new data.  This paper outlines work to generate
realistic synthetic data without those restrictions and for an important
domain:  credit card transactions -- including both normal and fraudulent
transactions.

At first glance it may appear simple to generate such transactions -- just
formalize a few items of the nature, ``Sally sold slacks to Sue on Sunday.''
However, there are many patterns and correlations in real purchases.  And
there are millions of merchants and innumerable locations.  And those
merchants offer a wide variety of goods.  Determining who shops where and
when becomes daunting.  Challenging also is the question of how much people
pay.  Inserting fraudulent transactions in the mix provides a final
challenge.

Addressing these many challenges and generating good data benefits from a
mixture of technical approaches and domain knowledge.  Those domains of
knowledge include mechanics of credit card processing as well as a broad set
of consumer domains, from electronics to clothing to hair styling to home
improvement and many more.  We also find that creation of a virtual world
depicting people's commercial lives facilitates generation of high-quality,
realistic data.  This paper outlines some of our key techniques and provides
evidence that the data generated is indeed realistic.

Although beyond the scope of this paper, our synthetic credit-card data also
facilitates development and training of models to predict fraud.  Those
models coupled with the synthetic dataset also provide foundations for
designing acceleration hardware, just as GPUs,
TPUs~\cite{NVidia_Volta17,TPU17} and other devices have been used for domains
such as image classification, object detection, natural lanaguage processing,
etc.

\end{abstract}

\section{Introduction}
\label{section-Intro}

As detailed next in Section~\ref{section-Credit_Card_Approach}, we use a
variety of techniques to synthesize credit card data and have implemented
them in a 40,000-line code base.  A key element of our approach is an
individual consumer so our data generation starts by creating models of
individuals.  We then create a population of individuals, with aggregate
characteristics mimic'ing their distribution in the real population.  Our
initial efforts are US-focused, so we broad characteristics are
representative of the United States, e.g. in age, occupation, income, credit
scores, geographic distribution, etc.

A first requirement of such statistics is that they match the mean and
standard deviation of the real population.  Generally census and other data
sources allow this requirement to be met in a straightforward, albeit
sometimes time-consuming manner.  Once we have means and standard deviations
for statistics of interest, we select specific values for individuals by
stochastic sampling, generally from a Gaussian distribution.

A fine point, but a key point here is the difference between the population
distribution and an individual's distribution.  For example, the population
(and different subgroups among it) have distributions for spending on certain
categories, e.g. spending on restaurant meals.  Once the mean and standard
deviation of an individual's restaurant spending is selected from the
population values, the individual spends according to their personal
distribution -- and does not redraw from the population numbers for each
purchase.  Without this distribution for the individual as opposed to the
population, an individual's spending would seem to fluctuate randomly with
high, medium, and low spending in proportion to the population.

However, there is a larger challenge than getting good values for mean and
standard deviation.  That challenge is obtaining accurate cross-correlations
between different metrics for an individual.  There are two primary reasons:

\begin{enumerate}

  \item Pairwise correlations are not available for every pair of statistics,
	e.g. haircare spending by US zipcode.  Even transitivity does not
	provide full or precise data, e.g. when the correlation between A and
	B is X, and between B and C is Y.  Some pairs of data have no
	transitive connection, but even when they do, combining X and Y
	generally yields a wide range for the correlation between A and C.

  \item Given generated data for two series A and B -- each with proper mean
	and standard deviation -- data for A and B must be transformed to
	create a specified correlation, while still maintaining the original
	means and standard deviations.  Thankfully, there are standard
	mathematical techniques using singular value decomposition which
	perform this operation~\cite{SVD-Correlation}.  Unfortunately these
	techniques are expensive both in memory and number of computations.
	Thus, we also employ more ad hoc techniques in some cases.  For
	example, foreign travel tends to increase with wealth.
	
\end{enumerate}

A population of individuals each with their own characteristics is a start.
However, actual behavior of those individuals must be instantiated.  For
example, what does a person buy?  When and where do they buy it?  How do they
buy it, e.g. cash, a particular credit card, etc?  We detail those apsects in
Section~\ref{section-Credit_Card_Approach}.

However, broadly speaking we simulate artificial worlds.  People live in
particular places, they travel for business and for pleasure, they don't work
most weekends, they buy things, etc.  To generate data, we log the
interactions people have in our simulated world.  Although we have not yet
taken broad advantage of the capability, our simulation approach enables
generation of heterogeneous data sets that are almost never available in real
data.  For example we could create unified datasets with credit card
transactions, loan applications, travel, callcenter conversations, and
medical data.  There are connections between all of these activities:  people
may purchase an item and then call to report a problem with it.  They may buy
airline tickets and then travel.  Medical visits incur expenses that paid by
credit or debit card.

As these interplays may suggest, another key component of our simulation
system is state machines.  For example, is a person in the {\sc Travel} state
or the {\sc Home} state.  The set of activities and purchases while at {\sc
Home} typically are different than in {\sc Travel}.  Similarly, purchases
tend to vary between {\sc Weekdays} and {\sc Weekends} and between {\sc
Morning}, {\sc Afternoon}, and {\sc Night}.  Our simulations use these states
and the transitions between them to generate more realistic data.  This state
machine approach couples well with our stochastic selection of values from a
distribution described above:  An individual's activities happen in natural
sequence.  However, the activities are not unrealistically mechanical.  For
example, chances are astronomically low that we would generate a person who
pays the same amount for dinner at the same restaurant at precisely the same
minute every Saturday evening.  However, a person may typically eat at
restaurants on Saturday evenings, with one restaurant being a particular
favorite.  Furthermore, there may be {\it concept drift}~\cite{Concept-Drift}
-- behaviors shifting over time.  Aside from the periodic shifts noted above
and the {\sc Home} / {\sc Travel} distinction, we model such things as
retirement and later, extreme age.

Another essential issue in generating data is accuracy and fidelity to real
behavior.  Our general approach to this problem is to compare synthetic data
to easily (and automatically) ascertained properties of real data.  For
example is synthesized spending on credit cards similar to real spending?  Do
people spend on the same sorts of things?  Are fraud levels similar?
Section~\ref{section-Results} provides more details and examples attesting to
the realism of our approach.

Once these automated checks have been done and the system is properly
configured, human experts can also be called upon to look at small samples of
synthetic data and check for any issues not detected by the automatic
assessment.  By the nature of its sampled approach, this human check can be
increased or decreased depending on the number of people available and the
budget for paying such people.  We emphasize that the measures (such as mean
spending per transaction or distribution of spending by merchant type)
employed in automatic comparisons do not require deep learning or human
labeled data.

Synthetic data has other benefits, e.g. it can be generated in different ways
so as to address particular challenges, e.g.

\begin{itemize}

  \item Improve explainability of results $\Rightarrow$ Generate data with a
	desired set of characteristics, e.g. a narrow range, bimodal values,
	etc.

  \item Avoid bias in results $\Rightarrow$ Generate data where two classes
	have statistically identical behaviors, and then check if the outputs
	of particular models are also statistically identical.

  \item ``Natural data'' is sparse or unbalanced $\Rightarrow$ Generate data
	that fills in the sparse areas or provides unrealistically high
	activity in a normally under-represented segment.

\end{itemize}

Our hardware colleagues have been industrious and successful in generating
chips and systems to accelerate training and inference, e.g. with
GPUs~\cite{NVidia_Volta17} or more specialized chips such as Google
TPUs~\cite{TPU17} or other chips~\cite{Habana-Training,Habana-Inference}.
However hardware has been developed and optimized primarily for learning
domains in which there is abundant labeled data, e.g.  image classfication,
object detection, speech recogntion, natural language processing,
translation, etc.  Learning domains such as credit card fraud that we target
with synthetic data do not have such accelerators.  Or at least hardware
designers did not have the performance of these domains as a primary design
consideration -- due to the relative lack of data and models.  Synthetic data
can change this situation and broaden the set of learning domains
participating in the virtuous circle of accuracy and performance
improvements.

\section{Synthesized Credit Card Data:  Approach}
\label{section-Credit_Card_Approach}

As noted in the Introduction, we generate synthetic credit card transactions
via simulation of a virtual world.~\footnote{For brevity we generally say
just ``credit cards'', but unless otherwise indicated that designation also
includes debit cards and prepaid cards.} That virtual world has a population
of consumers with characteristics such as age, income, and geographic
location in proportion to the overall population of the United States.
(Eventually we hope to extend our model to have consumers across the world.)

The virtual world also has a population of merchants.  Like consumers,
merchants embody many real-world characteristics.  For example we model total
sales amount by merchant category.  For credit cards, categories are labeled
by a Merchant Category Code or MCC.  MCC codes range from 0 - 9999.)  We
model both large multinational merchants like Apple (MCC=5045) and McDonalds
(MCC=5814) as well as local merchants such as dry cleaners (MCC=7210).  The
model has over 300 multinationals, each with many physical locations and in
most cases an online presence.  Unlike consumers, merchants are not limited
to the U.S, but are distributed around the world.  (Consumers based in the
U.S. may travel anywhere in the world, and a key component of fraud detection
is separating actual consumer travel and consequent purchases from fraudulent
activity.)  Altogether our model has over 16 million merchant locations at
which consumers can shop.

Unfortunately a representation that includes only consumers and merchants
makes it challenging to determine specific stores where transactions should
occur.  Consumers often shop for specific items, e.g. clothing or groceries
or a necklace.  Those items and most others can be purchased at a broad set
of MCCs -- from stores like Walmart (MCC=5311) selling a wide range of
merchandise to stores selling a narrow range of items like a local jeweler
(MCC=5094).  Thus, internal to our virtual world we have created a list $GS$
of almost 100 types of {\underline {\it G}}oods and {\underline {\it
S}}ervices that people purchase, and a mapping from $GS$ to the set of MCCs
at which the goods and services can be purchased.  The MCCs then map to
merchants and specific merchant locations.

$GS$ also has other features, e.g.  how frequently an item is purchased.
(This frequency is actually a distribution ranging from the fraction of
people who buy the item multiple times per day to the fraction that buy the
only once a decade.)  $GS$ also captures time-of-day tendencies, e.g.  that
people are more likely to visit a bar in the evening than the morning.
Similarly $GS$ notes the relative proclivity to consume an item on weekdays
vs weekends and whether at home, on vacation, or on business travel.  Finally
each item in $GS$ has an income distribution -- indicating both the
likelihood that people buy the item and the amount they spend on it if they
do buy it.  These $GS$ characteristics are then translated into specific (and
different) preferences for each consumer.

However, just knowing tendencies and preferences does not adequately capture
consumer behavior.  As noted in the Introduction, our virtual world includes
state machines to reflect causal relationships, for example making purchases
at merchants relatively close to the consumer's current location.

\section{Synthesized Credit Card Data:  Examples}
\label{section-Credit_Card_Examples}

\begin{figure}
\begin{center}
\includegraphics[width=2.5in,height=1.8in]{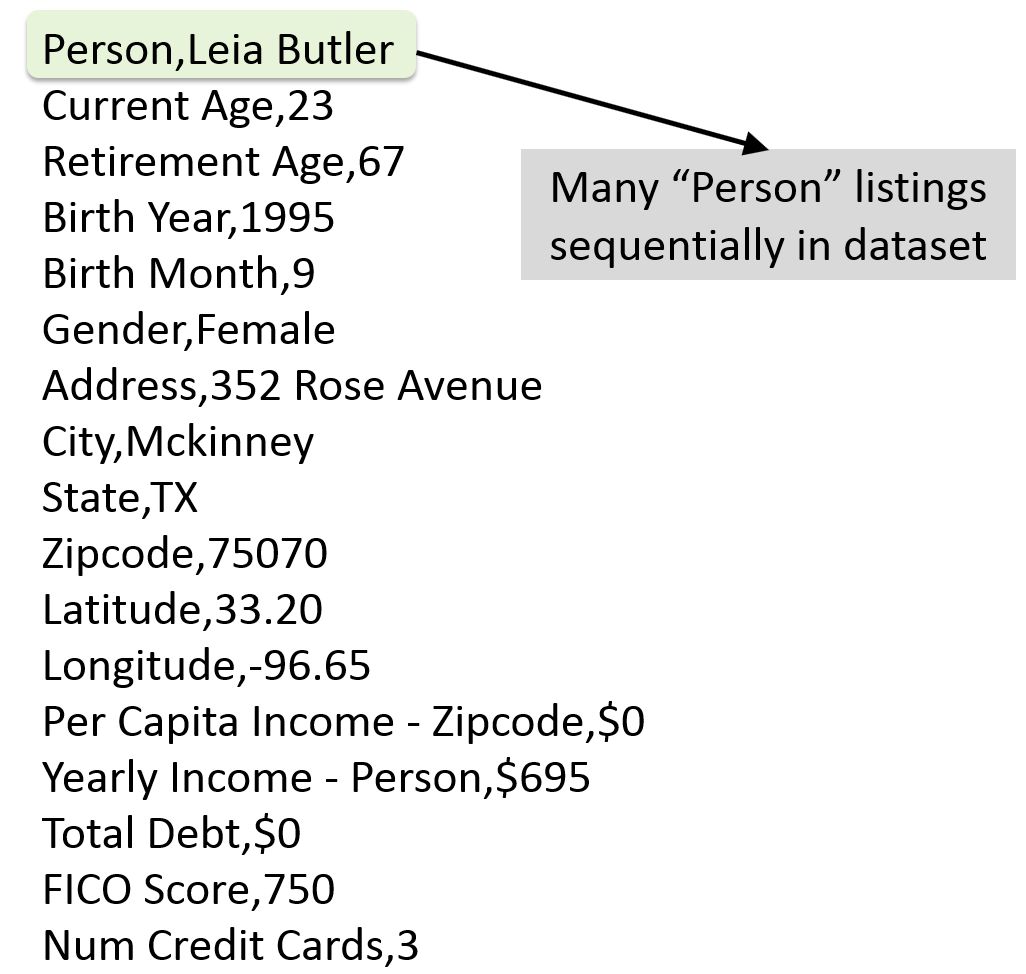}
\end{center}
\caption{Sample Bio of Consumer, ``Leia Butler''}
\label{Fig-Credit_Card_Consumer_Bio}
\end{figure}

\begin{figure}[t]
\includegraphics[width=1.5in]{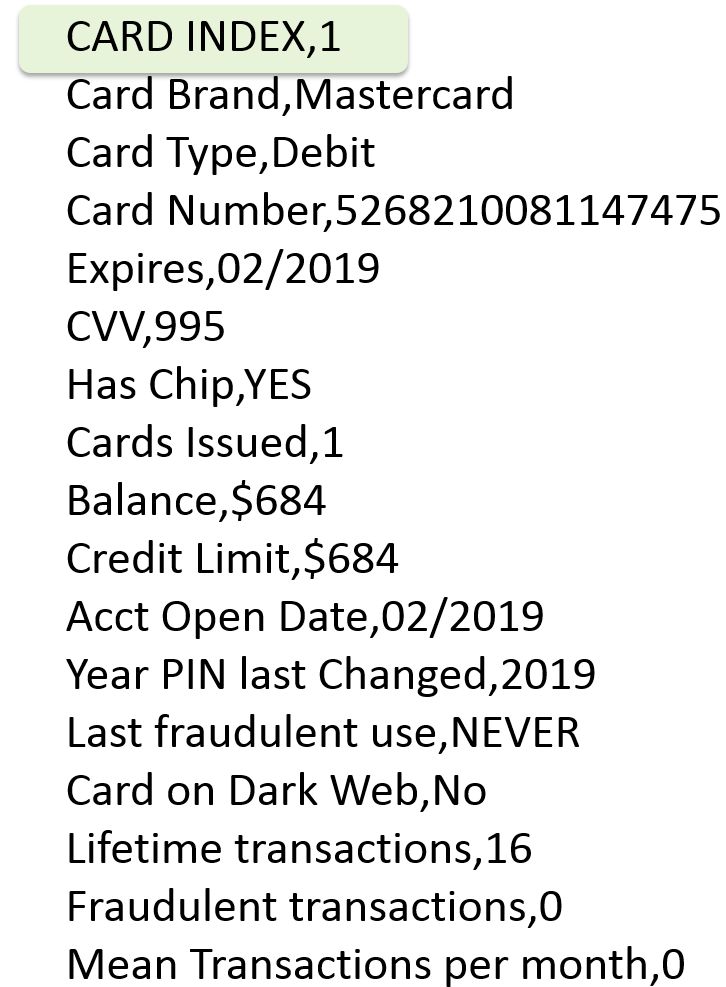}
\caption{Sample Info: Card of Consumer, ``Leia Butler''}
\label{Fig-Credit_Card_Info}
\end{figure}

\begin{figure}
\begin{center}
\includegraphics[width=\linewidth]{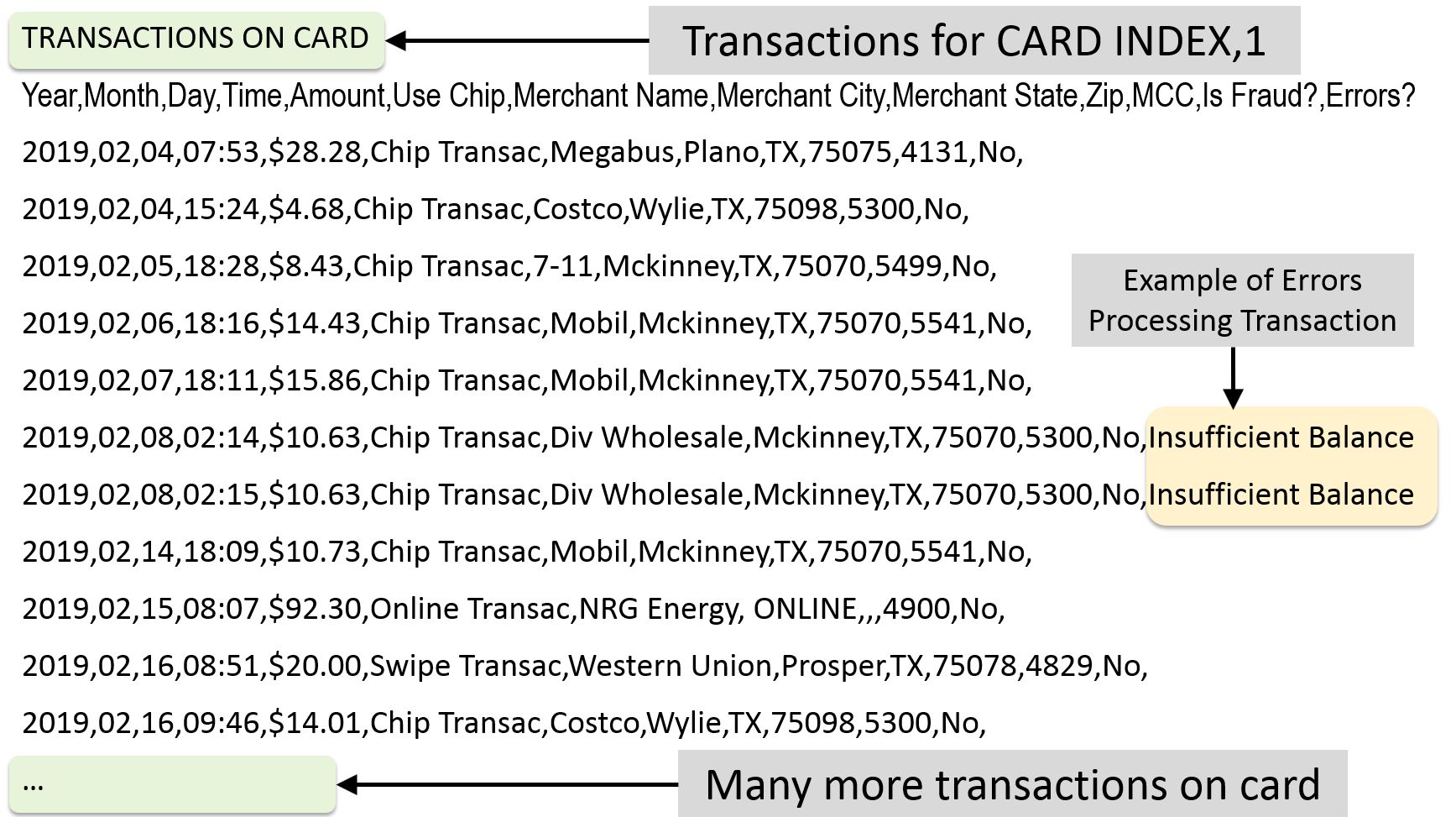}
\end{center}
\caption{Sample: Transactions of Consumer, ``Leia Butler''}
\label{Fig-Credit_Card_Transactions}
\end{figure}

\begin{table}[t]

\noindent\makebox[\linewidth]{\rule{\linewidth}{0.4pt}}

\begin{itemize}

  \item Mean and Standard Deviation for Cards per Consumer, Cards per
	Account, Transactions per Year, FICO Score, Income, Debt as fraction
	of income, Credit Limit, Balance as fraction of Credit Limit, Years
	Account Open, Years since last PIN change

  \item Mean Annual Weekend Getaways
  \item Mean Annual Vacations
  \item Mean Vacation Duration
  \item Mean Annual Business Trips
  \item Mean Business Trip Duration
  \item Probability of Foreign Weekend Getaway
  \item Probability of Foreign Vacation
  \item Probability of Foreign Business Trip

\end{itemize}
\noindent\makebox[\linewidth]{\rule{\linewidth}{0.4pt}}

\caption{Selected parameters controlling credit virtual world}
\label{Tbl-Parms_for_Virt_World}
\end{table}

To make these notions concrete, Figure~\ref{Fig-Credit_Card_Consumer_Bio}
shows sample biographical data of one consumer (``Leia Butler'').  The bottom
of Figure~\ref{Fig-Credit_Card_Consumer_Bio} notes that Leia has 3 cards.
Figure~\ref{Fig-Credit_Card_Info} then shows one of the 3 cards (a debit
card) generated for Leia.  Finally, Figure~\ref{Fig-Credit_Card_Transactions}
shows a sampling of transactions on that card.  These transactions are
generated using the stochastic sampling techniques and state machines
described above, and with the parameters in
Table~\ref{Tbl-Parms_for_Virt_World} as some of the key inputs.

\noindent {\bf Synthesizing Fraud}

A key purpose of generating synthetic credit card data is to help train
models to do a better job of detecting fraud.  As such, the virtual world
must include not only genuine transactions between consumers and merchants,
but also fraudulent transactions.  We have implemented two mechanisms for
fraud:

	The first mechanism creates a population of fraudsters similar to the
	population of consumers.  Each fraudster lives in a particular place,
	has particular preferences for items purchased, days of the week when
	purchases are made, etc.  Each fraudster is also active for a
	particular time range -- from months to years.  This model comports
	with real observations on fraudster
	behavior~\cite{Fraudster-Behavior}.  It also reflects that for most
	fraudsters, using cards for false purchases is not a hobby but their
	job.  Like other jobs, fraud is carried out on a particular schedule
	and in particular places.  And like other jobs, workers enter and
	leave the field.

	The second mechanism generates fraudulent transactions at random
	points in time for each consumer.  This {\it random} mechanism could
	represent a worst-case future scenario when fraudsters have
	determined how to randomize their purchases among stolen cards so
	that there is little apparent pattern to their purchases.  This case
	represents another benefit of synthetic data:  the potential to get
	ahead of the curve of real data, and to determine how well models
	work in hypothetical what-if scenarios.

With a population of fraudsters as in the first mechanism, we also label the
generated transactions with the identity of the fraudster.  Using this label
for training a model is of course forbidden:  real transactions never come
with such labels.  However, fraudster labels have two benefits:

\begin{itemize}

  \item The labels can ease debugging and understanding of models.  If a
	model does particularly well or poorly identifying fraud from a
	particular fraudster, that info can be used to further tweak the
	model and improve its accuracy.

  \item Our generation of synthetic transactions is independent of the model
	for detecting fraud.  Thus, during data generation it is not known
	how quickly a model will detect fraudulent transactions and revoke
	the card to stop the fraud.  Once a model detects fraud with from
	fraudster $F$ on card $C$, the model can throw away future $F$
	transactions on card $C$.  This capability is not available when
	training with real data.  If the deployed fraud-detection model
	detects fraud on $C$ at a particular time, there will be no future
	fraudulent transactions on $C$.  Thus, when models are trained on
	real data, they become dependent on the behavior of previous models.

\end{itemize}

\noindent {\bf Synthesizing Other Attributes}

Another important element of our virtual credit card world is modeling the
chip / non-chip status of credit cards and debit cards.  Chips generating
unique transaction identifiers were introduced on a large scale in the U.S.
in 2014.  Compared to the previous magnetic stripe technology, the chip's
unique identifiers make it harder to perpetrate ``card-present'' fraud.  As a
result of (1) this chip technology; (2) increasing numbers of online
transactions; and (3) increasing thefts of credit card information from large
online repositories -- online purchases now dominate credit card fraud.
Approximately 70\% of fraud in the U.S. now happens
online~\cite{Fed-Fraud-Study}.  Europe adopted chip cards earlier and saw
online fraud increase commensurately sooner.

Our model can generate consumers over an arbitrary period of time.  We
typically start in the mid 1980s and simulate until the present.  We model a
number of the changes over this span.  For example, online transactions start
in the mid 1990s and gradually grow to present levels.  As just noted, levels
of online fraud also increase significantly in the last few years.  Over this
long time period, 18-year old consumers and others also emerge and begin
using cards for the first time.  Others retire and their pattern of purchases
change.

Data from such long time periods is unavailable in the small number of real
data sets on which work has been
published~\cite{Pozzolo14,Pozzolo15,Carcillo18} -- yielding another
benefit of synthetic data.  Reports about fraud detection using real data
have periods ranging from 2 days to a few months to a one year.  However,
many purchases typical of fraud occur at long time intervals.  For example,
foreign trips are separated by years for many people.  Expensive purchases
such as furniture, jewelry, and high-end electronics also tend to be
purchased relatively infrequently.  Like travel, these purchases are
disproportionately represented in fraud.  Thus data spanning long time
periods is important to separating real transactions from fraud.

Our synthetic data has {\it all} credit and debit cards of a consumer as well
as their cash purchases.  Real data sets are typically limited to
transactions from one card or at most one family of cards (e.g. Visa or
Mastercard) and never include cash. As such, synthetic data provides a means
for assessing how much accuracy is lost due to unavailability of ``full''
data sets.  A broad set of synthetic data also provides a foundation for
transfer learning and augmenting real data (as opposed to totally supplanting
it as we do).

\section{Results}
\label{section-Results}

\begin{table}[t]
\noindent\makebox[\linewidth]{\rule{\linewidth}{0.4pt}}
\begin{center}
\includegraphics[width=\linewidth]{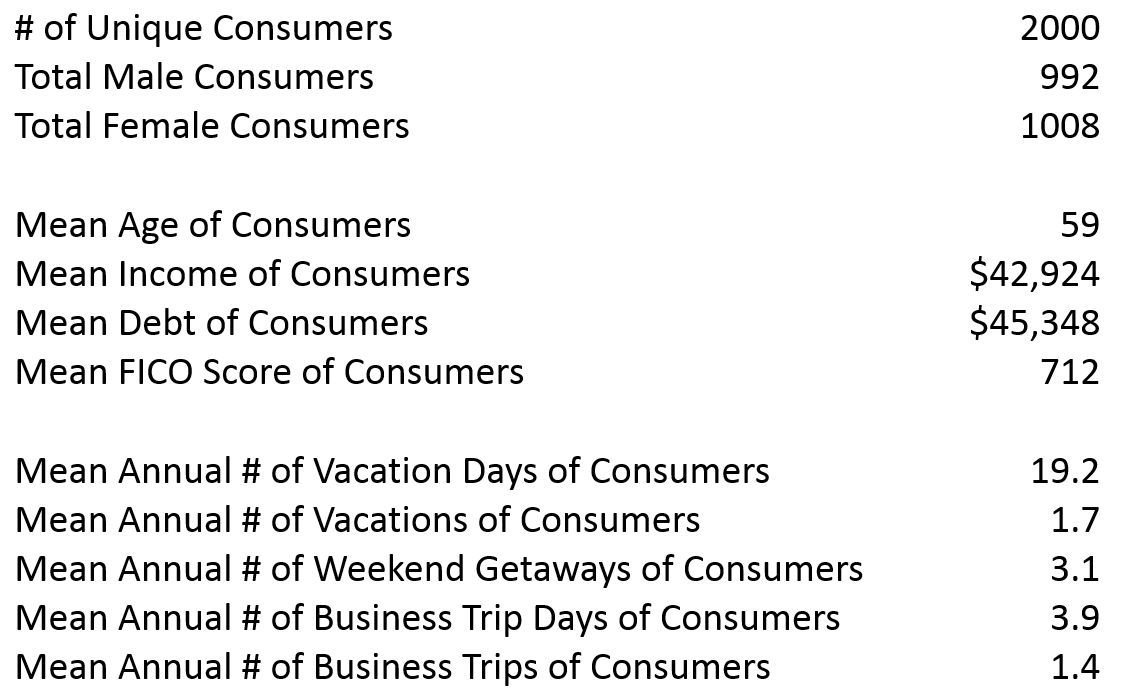}
\end{center}
\noindent\makebox[\linewidth]{\rule{\linewidth}{0.4pt}}
\caption{Summary stats of biographical attributes}
\label{Tbl-Credit_Card_Consumer_Stats}
\end{table}

To fine-tune data generation and provide fidelity with the real world, we
look at many population statistics.
Table~\ref{Tbl-Credit_Card_Consumer_Stats} summarizes across the biographical
attributes listed in Figure~\ref{Fig-Credit_Card_Consumer_Bio}.  If the
summary stats do not match what is desired, we can adjust the values in
Table~\ref{Tbl-Parms_for_Virt_World} and others until the population
aggregates have the desired values.

\begin{table}[h]
\noindent\makebox[\linewidth]{\rule{\linewidth}{0.4pt}}
\begin{center}
\includegraphics[height=2.0in]{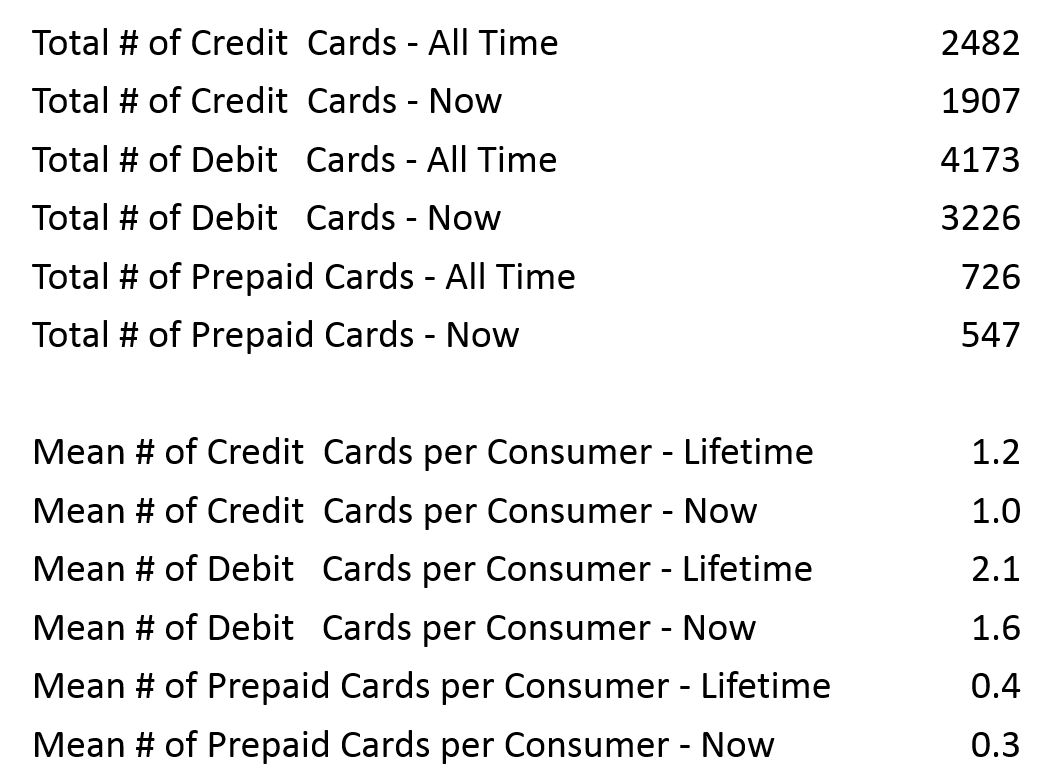}
\end{center}
\noindent\makebox[\linewidth]{\rule{\linewidth}{0.4pt}}
\caption{Summary stats of credit, debit and prepaid cards}
\label{Tbl-Credit_Card_Stats}
\end{table}

Similarly and as analog to Figure~\ref{Fig-Credit_Card_Info},
Table~\ref{Tbl-Credit_Card_Stats} provides a summary across the credit,
debit, and prepaid cards of individuals.  The numbers in
Tables~\ref{Tbl-Credit_Card_Consumer_Stats} and~\ref{Tbl-Credit_Card_Stats}
are indeed reflective of the broad U.S. population, e.g. roughly equal
numbers of men and women, mean FICO score of 712, mean income, vacation days,
etc.

\begin{table}[t]
\begin{center}

  \begin{subtable}{\linewidth}
  \includegraphics[width=\linewidth]{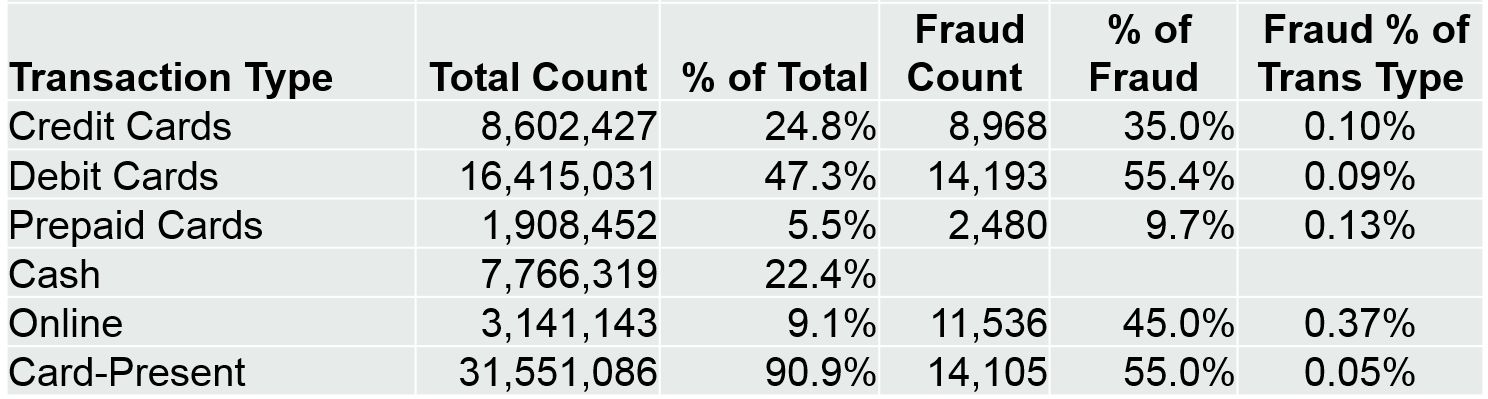}
  \caption{Lifetime Statistics on Transactions}
  \end{subtable}

  \begin{subtable}{2.6in}
  \begin{center}
  \includegraphics[height=1.1in]{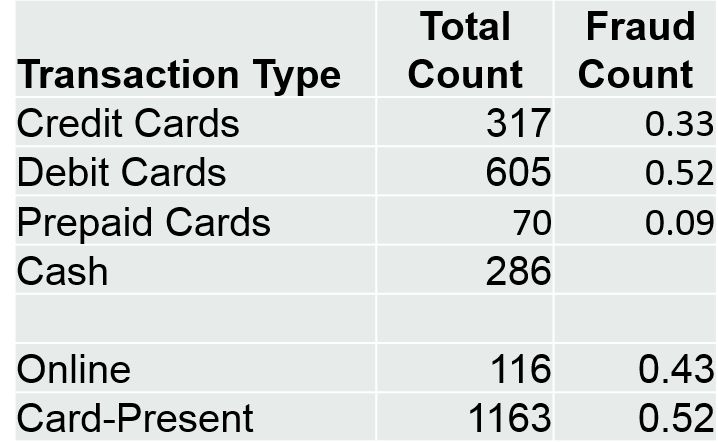}
  \end{center}
  \caption{Annual Statistics on Transactions}
  \end{subtable}

  \begin{subtable}{\linewidth}
  \begin{center}
  \includegraphics[height=1.0in]{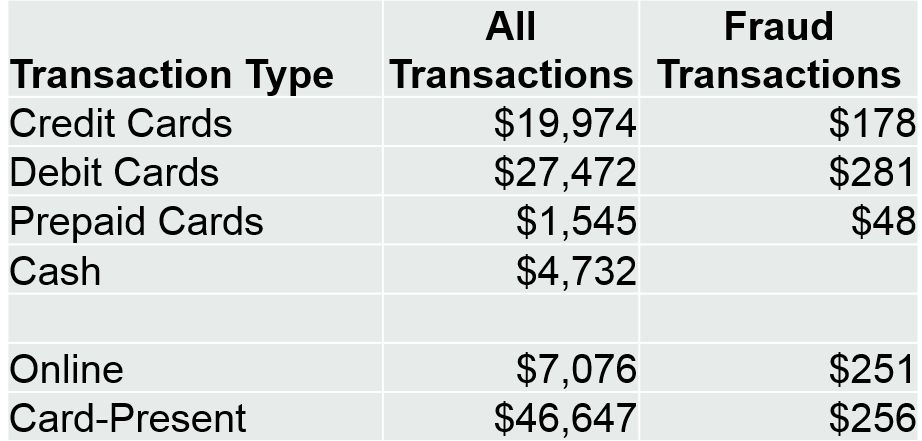}
  \end{center}
  \caption{Annual Statistics on Spending}
  \end{subtable}

  \begin{subtable}{\linewidth}
  \begin{center}
  \includegraphics[height=1.0in]{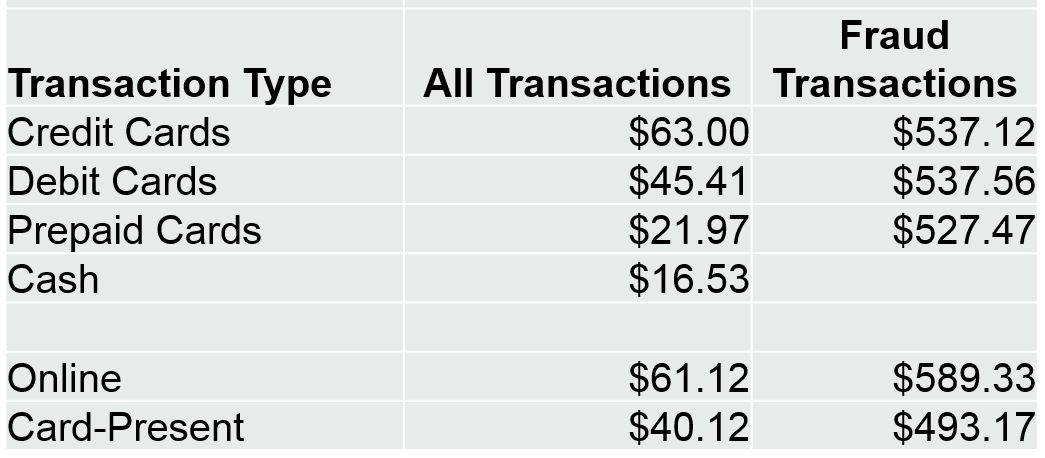}
  \end{center}
  \caption{Statistics on Spending per Transaction}
  \end{subtable}

\end{center}
\caption{Summary stats:  transactions and spending}
\label{Tbl-Credit_Transaction_Stats}
\end{table}

It is also useful to look at summary statistics on transactions at varying
granularities.  In particular we look at 
(a) lifetime transactions -- over the approximately 35 years of the virtual
world in 
Table~\ref{Tbl-Credit_Transaction_Stats}(a); 
(b) annual transaction counts in 
Table~\ref{Tbl-Credit_Transaction_Stats}(b);
(c) annual spending in
Table~\ref{Tbl-Credit_Transaction_Stats}(c); and 
(d) per-transaction spending in
Table~\ref{Tbl-Credit_Transaction_Stats}(d).  

The numbers in these tables reflect actual values, e.g. the amounts of
different transaction types in Table~\ref{Tbl-Credit_Transaction_Stats}(d) --
for both fraud and non-fraud.  Similarly the usage ratio of credit, debit,
and prepaid cards in Table~\ref{Tbl-Credit_Transaction_Stats}(a) is accurate,
as is the higher rate for online fraud than card-present. We also hope that
readers will find the transactions in
Figure~\ref{Fig-Credit_Card_Transactions} realistic.

We tabulate many other statistics beyond the data shown in these tables, e.g.
statistics per U.S. state, per country, and per MCC, as well as various
online statistics.  We omit details here for brevity.

Once input parameters are such that summary statistics for the generated data
match desired values, we can generate arbitrarily large datasets.  We have
generated datasets spanning 35 years with 20,000 consumers performing more
than 300 million transactions.  Represented in CSV format, such a dataset
requires over 20 GBytes.  Arbitrarily larger datasets are possible as
needs dictate and storage resources allow.

\section{Related Work}
\label{section-RelWork}

Many previous works assess fraud-detection
models~\cite{Kaggle-Model-1,Kaggle-Model-2}.  The largest number have been
built around a public-domain Kaggle dataset~\cite{Kaggle18} with about
280,000 transactions collected over 2 days.  Our transaction count is more
than 1000x larger and spans decades, not days.  Unlike our data the Kaggle
data also obfuscates all features except transaction dollar amount and time.
This obfuscation is done via a principal components transformation which
creates uncorrelated numerical features.  This lack of correlation is
unrealistic.  Having no underlying intuitive sense of the features also
increases the difficulty of building models.  Capital One has blogged about
their internal work using GANs to generate synthetic transaction
data~\cite{Capital-One-Synth-Data}.  However their approach requires access
to real data, which is then amplified to create new data.  Our approach
requires no access to real transactions.

Other studies have been done on more realistic (non-public) data
sets~\cite{Pozzolo14,Pozzolo15,Carcillo18}.  However, the time span for these
studies ranges from 3-12 months versus 35 years for our data.  The maximum
number of transactions in these previous studies is around 10 million -- less
than 1/30-th of the number in our synthetic data, and we can generate
datasets that are far larger still.  As has been observed in other domains,
the quantity of data matters in achieving high model accuracy.

More broadly synthetic data can be viewed as a complement to techniques such
as transfer learning~\cite{Pratt93} and few-shot
learning~\cite{Murphy03,Bart05} that learn based on a small set of data.
However, instead of having models learn from a small amount of data or from
results in a related domain, our approach generates data from fundamental
principles of how things work and simulations embodying those principles.

\section{Conclusions}
\label{section-Conclusion}

We have outlined techniques for synthesizing credit card transactions for
U.S.-based consumers purchasing in the U.S. and world-wide.  We have
also provided statistics and transaction snippets indicating that the results
are realistic.

Improvements are always possible.  As future work we plan to support broader
populations than the U.S.  We also plan to enhance the state model in our
virtual world to provide yet more realism in individual consumer behavior.
We also plan to examine how GANs~\cite{Goodfellow14} could systemically
improve our data.

Further afield, many techniques outlined here can be applied to sythesize
other types of data.  Bank loan applications have many overlaps as do patient
medical records.  Medical records have at least as many privacy restrictions
as credit card data and can also benefit from a virtual world approach for
modeling behavior, disease progression, etc.

Transcripts are available for a large corpus of speech.  However, except in a
few cases involving large human effort, transcripts do not provide the
underlying semantic intent of the words.  Synthetic approaches could prove
helpful.  Automatically interpreting charts and graphs is another challenge
where synthetic data may help.

Aside from new domains and improvements in real-world fidelity, we plan to
investigate improved technical approaches.  For example with thousands of
variables, generating cross-correlations between all pairs is computationally
expensive.  Can we improve it?

\section{Acknowledgements}
\label{section-Ack}

In the course of numerous conversations my colleagues at IBM have provided
much useful feedback and insight.  In particular I thank Shyam Ramji, Ravi
Nair, and Jeetu Raj.


\begin{small}

\bibliographystyle{aaai}
\bibliography{AAAI20_Bib_v2_short}

\end{small}

\end{document}